\documentclass[aps,prd,superscriptaddress,nofootinbib,showpacs,twocolumn]{revtex4-1}\pdfoutput=1
\usepackage{amsmath,latexsym,amssymb,graphicx,ifpdf}
\usepackage{hyperref}
\usepackage{url}
\hypersetup{colorlinks,citecolor= nicegreen,linkcolor= nicered}
\usepackage{color}
\usepackage{cancel}
\definecolor{red}{rgb}{1,0,0}
\definecolor{nicered}{rgb}{0.7,0.1,0.1}
\definecolor{nicegreen}{rgb}{0.1,0.5,0.1}

\newcommand\GeV{\text{GeV}}
\newcommand\TeV{\text{TeV}}

\newcommand{\beq}{\begin{equation}}
\newcommand{\eeq}{\end{equation}}
\newcommand{\bea}{\begin{eqnarray}}
\newcommand{\eea}{\end{eqnarray}}

\begin{document}

%
%
%

\title{Inert Doublet Dark Matter and Mirror/Extra Families after Xenon100}

\author{Alejandra Melfo}
\affiliation{Universidad de Los Andes, M\'erida, Venezuela}
\affiliation{ICTP, Trieste, Italy}
\author{Miha Nemev\v sek}
\affiliation{ICTP, Trieste, Italy}
\affiliation{J.\ Stefan Institute, Ljubljana, Slovenia}
\author{Fabrizio Nesti}
\affiliation{ICTP, Trieste, Italy}
\author{Goran Senjanovi\'c}
\affiliation{ICTP, Trieste, Italy}
\author{Yue Zhang}
\email{amelfo@ictp.it, miha@ictp.it, goran@ictp.it, nesti@aquila.infn.it, yuezhang@ictp.it}
\affiliation{ICTP, Trieste, Italy}
\date{\today}

\begin{abstract}%
\noindent It was shown recently that mirror fermions, naturally present in a number of directions for new physics, seem to require an inert scalar doublet in order to pass the electroweak precision tests. This provides a further motivation for considering the inert doublet as a dark matter candidate.  Moreover, the presence of extra families enhances the Standard Model Higgs-nucleon coupling, which has crucial impact on the Higgs and dark matter searches. We study the limits on the inert dark matter mass in view of recent Xenon100 data. We find that the mass of the inert dark matter must lie in a very narrow window $75\pm1$\,GeV while the Higgs must weigh more than 400\,GeV. For the sake of completeness we discuss the cases with fewer extra families, where the possibility of a light Higgs boson opens up, enlarging the dark matter mass window to $\frac{1}{2}m_h$--$76\,$GeV.
%
%
We find that Xenon100 constrains the DM--Higgs interaction, which in turn implies a lower bound on the monochromatic gamma-ray flux from DM annihilation in the galactic halo. For the mirror case, the predicted annihilation cross section lies a factor of 4--5 below the current limit set by Fermi LAT, thus providing a promising indirect detection signal.
\end{abstract}



\pacs{14.65.Jk, 14.60.Hi, 14.80.Ec, 95.35.+d}

\maketitle

\section{Introduction}

\noindent Recent data released by the Xenon collaboration~\cite{Aprile:2011hi}, relative to 48000\,kgd statistics, improve previous limits on the Dark Matter (DM) direct detection cross section versus the DM mass. This result has significant implications for scenarios of Dark Matter based on particle physics.

One popular such scenario, pursued in recent years, is the inert scalar doublet extension of the minimal Standard Model~\cite{Barbieri:2006dq}. While its great virtue lies in its simplicity, the stability of the inert candidate is assumed without any theoretical hint in its favor.

As we argued in a recent paper~\cite{Martinez:2011ua}, the inert nature of the second doublet is favored by electroweak precision tests (EWPT) constraints in the presence~\cite{He:2001tp} of mirror families.  Mirror fermions are a must in a number of physically motivated scenarios: Kaluza-Klein theories~\cite{Witten:1981me}, family unification based on large orthogonal groups~\cite{GellMann:1980vs,Senjanovic:1984rw,Bagger:1984rk}, N=2 supersymmetry~\cite{delAguila:1984qs} and some unified models of gravity~\cite{GGUT}. Moreover, they were envisioned by Lee and Yang in their classic paper on parity breakdown~\cite{Lee:1956qn} as a way to restore parity in the fundamental interactions.

Such a framework becomes quite predictive when constraints from the EWPT and the vacuum stability of the Higgs potential~\cite{Martinez:2011ua} are considered.  The lowest component of the inert doublet, which is a possible dark matter candidate, must have a mass less than around $100\,$GeV.

In view of this, we study the dark matter direct detection in the presence of three extra chiral mirror families, taking into account the recent data from the Xenon100 experiment.  The key point here is the enhancement of the effective Higgs portal to the nucleon in the presence of the extra heavy fermions~\cite{Shifman:1978zn}. This is precisely the same diagram which enhances the Higgs production cross section at hadron colliders. In the presence of mirror families the enhancement of direct detection gives approximately a factor of 9, so with the new Xenon100 bound this scenario becomes predictive for the mass and possible annihilation channels of the inert dark matter.
In turn, this frameworks predicts a lower bound on the monochromatic gamma ray line from the annihilation in the galactic halo, whose cross section is less than an order of magnitude below the current Fermi LAT sensitivity.
%
%


It is important to keep in mind that although the mirror case provides a good rationale for the stability of the inert DM candidate, generically it may not be necessary to stick to the mirror conjecture. In fact, none of the results we present depend on the choice of chirality of the extra fermions, and as such they are equally applicable to the usual additional copies of the SM families. In what follows, we present our results for one to three extra families (more are not allowed by precision tests) in the presence of an extra inert doublet.

The paper is organized as follows. In the section II, we describe the theoretical and experimental constraints on the inert doublet extension of the SM with extra families. We include the limits from direct collider searches, the electroweak precision constraints, perturbativity and vacuum stability, and describe the updated Tevatron Higgs exclusion window in this model.  In Section III, we present our results for the relic density and direct detection, which constrain the DM mass to lie between $\frac{1}{2}m_h$--$76\,$GeV for a fourth family and in a very narrow window $75 \pm 1$\,GeV in the case of the three mirror families. This enables us to give a quite robust prediction for the monochromatic gamma ray flux from the galactic halo. Section IV contains a summary and the outlook.

\pagebreak[3]

\section{Extra/Mirror Families Seeking Friendship with an Inert Doublet}

\noindent 
As we will describe now, the EWPT together with the constraints from perturbativity and vacuum stability, favor the existence and inertness of an extra doublet, $\Phi$.  After the electroweak symmetry breaking, this field decomposes into an extra scalar $S$, a pseudoscalar $A$ and a charged component $C$.

\subsection{Vacuum stability and perturbativity}

\begin{figure}[b]
\vspace{-2ex}%
\centerline{\includegraphics[width=.9\columnwidth]{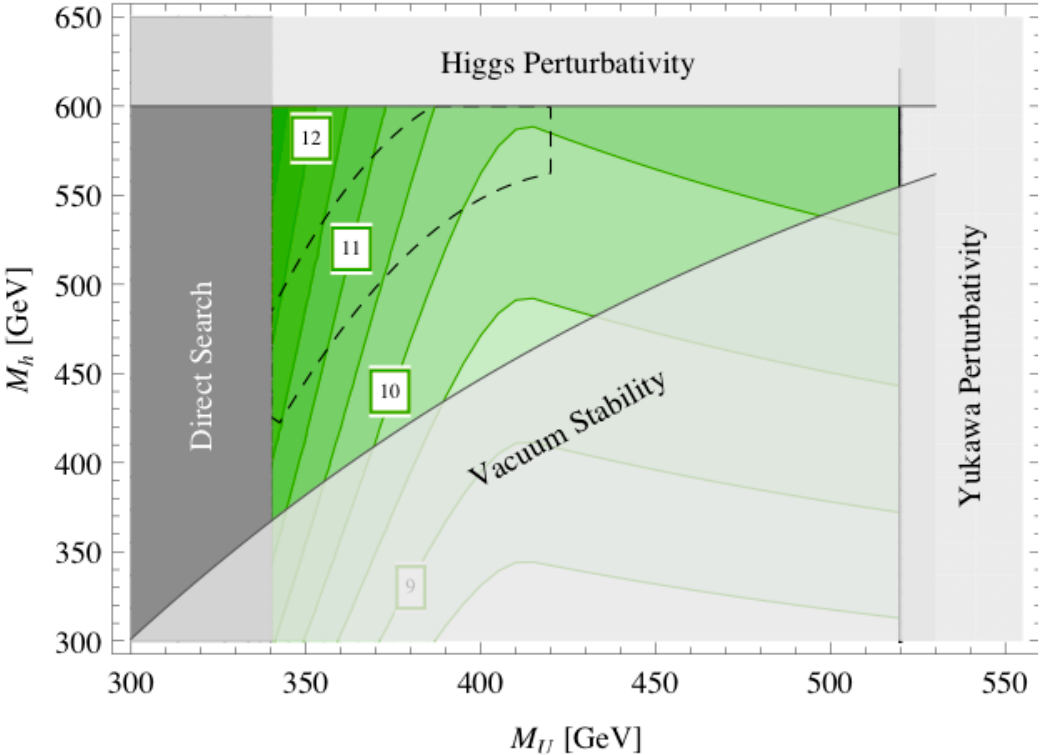}}%
\vspace{-2ex}%
\caption{Stability and perturbativity limits on $m_h$, $m_U$ (shaded gray regions), in the presence of three extra generations. The central allowed region corresponds to a low cut-off $\Lambda=700\,\GeV$; the dashed-contour region to $\Lambda=1\,\TeV$. In the background, we report the $\chi^2$ contours from Fig.~\ref{chi2}, fourth panel, showing that the best points, with reasonably high cutoff, lie around $m_U\sim 400\,\GeV$ and $m_h\sim 500\,\GeV$.\label{stability}}%
\vspace{-1.ex}
\end{figure}

The large Yukawa couplings of the extra quarks become a problem for vacuum stability in the case of a light SM Higgs, as they tend to drive the Higgs boson self coupling to negative values. Therefore, one should take the extra quarks as light as possible, without running into conflict with direct search, $\sim 350\,$GeV.  Still, in the case of more than one extra family, the light Higgs mass window $115\,\GeV \lesssim m_h \lesssim 131\,\GeV$ is excluded if the vacuum is to be stable up to a reasonably high cutoff (e.g.\ 1\,TeV).  In particular with three extra families (mirror case), the SM Higgs boson needs to be heavier than $\sim400\,\GeV$~\cite{Martinez:2011ua}.  At the same time the requirement of perturbativity imposes an upper bound on the SM Higgs mass, which is about 600\,GeV~ \cite{Barbieri:2006dq, LopezHonorez:2006gr}, as well as on the mass (Yukawa couplings) of heavy fermions, which have to be lighter than roughly 500\,GeV.  All this has important implications for the discussions of the EWPT, in the next section (see Fig.~\ref{stability}).

Concerning the components of the extra doublet, in the mass ranges that are favored by EWPT, they have no appreciable impact on perturbativity or/and stability of the SM Higgs.

\subsection{Electroweak Oblique Corrections}

It has been shown in~\cite{He:2001tp, Flacher:2008zq} that with a fourth family one can fit the electroweak oblique parameters $S$, $T$, $U$ to within 68\% confidence level (CL). However, this becomes progressively constrained as more families are added, until $\chi^2 \approx 13.5$ for the case with three extra families, outside 99\% CL.

The introduction of the second doublet can help to alleviate the tension (in appendix~\ref{app} we list the relevant expressions for its contribution).  In fact, we have explored the best fit cases and find that they are characterized by a significant cancellation of the contributions to $T$ from the (three) extra families and the doublet.  In this case we find the best $\chi^2 \approx 9.0$, lying inside the 99\% ellipse.

Qualitatively, one can understand what happens for the mirror case in the following way:
Naively for heavy electroweak doublet fermions, the contribution to the $S$ parameter is $1/(6\pi)$. Therefore, three extra families contribute around $0.7$. However, as noticed recently~\cite{Murayama:2010xb}, this relatively large contribution can be reduced by making the extra neutrinos lighter than the $Z$-boson.
The $T$ parameter receives a large positive contribution by splitting the extra neutrinos and charged leptons, which can be compensated by splitting the components of the second Higgs doublet.
Finally, the reason one cannot get even an even better fit is due to the relatively large contribution in the $U$ direction, again from splitting the lepton doublets of the extra families.

\begin{figure}[b]
\vspace*{-3.5ex}%
  \centerline{\includegraphics[width=0.9\columnwidth,height=.65\columnwidth]{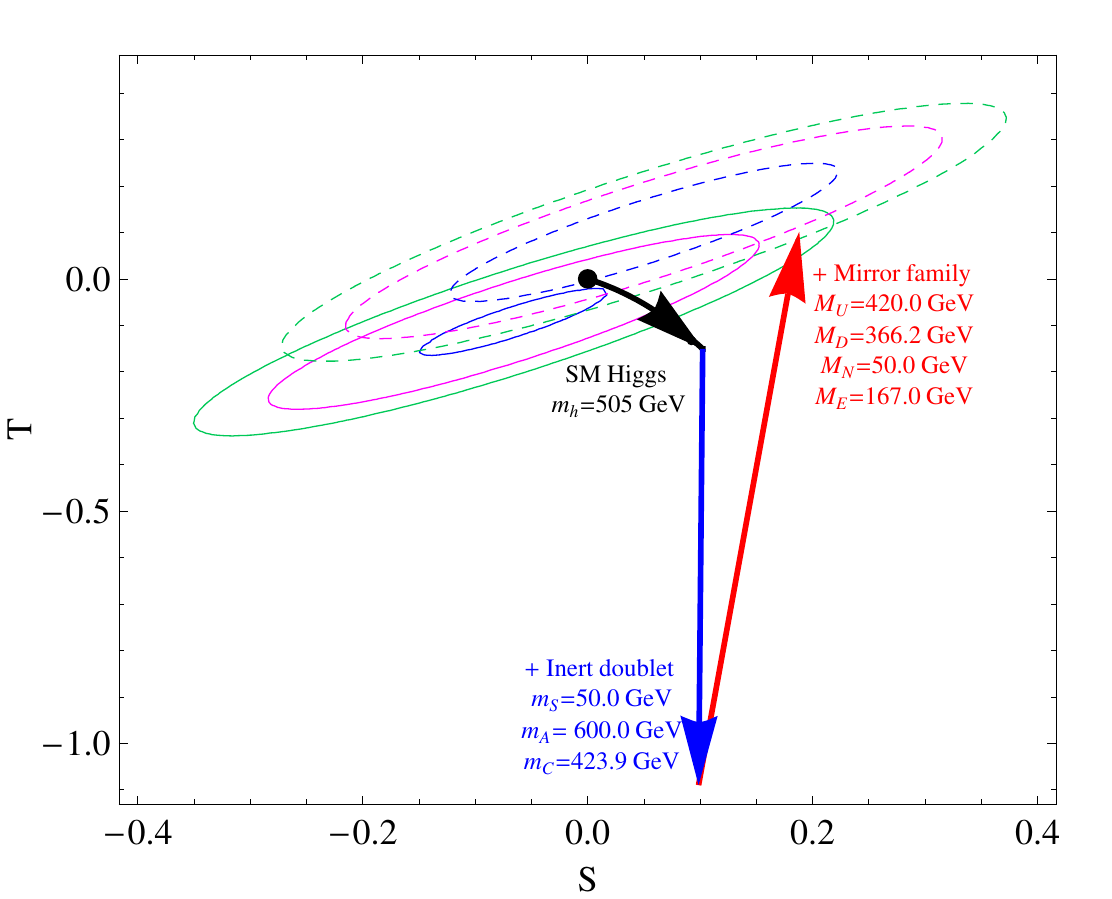}}%
\vspace*{-1ex}
  \caption{An illustration of best fit point for the electroweak oblique corrections (the $S-U$ plane) with inert Higgs doublet and extra families.  The contributions from SM Higgs, second doublet and mirror fermions are added in order. The best fit is associated with a large cancellation in the $T$ direction. The arrows starts from the reference plane where $U=0$ (dashed) and end up on the plane with $U=0.269$ (continuous). \label{obliquehih}}%
\end{figure}

\begin{figure*}[t]
\vspace{-2ex}%
\def\hh{.49}
\includegraphics[width=\hh\columnwidth]{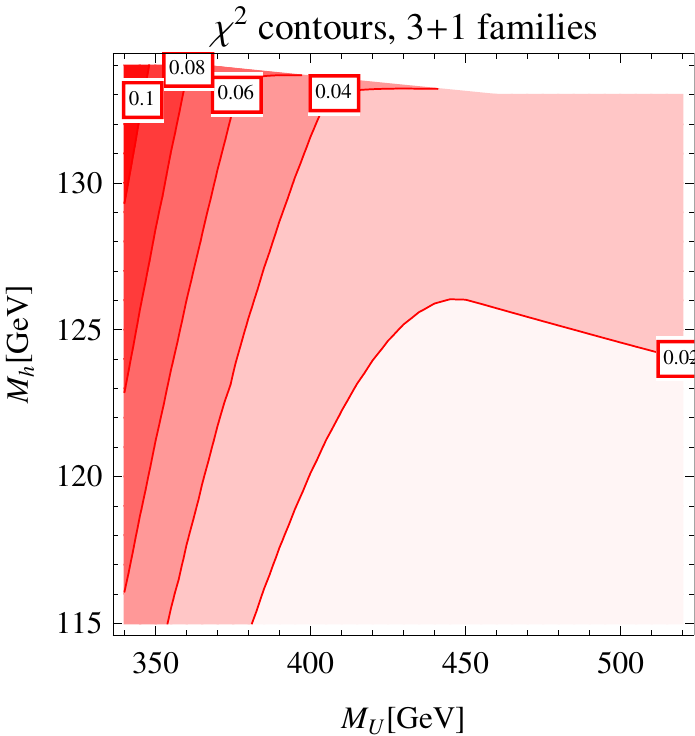}~
\includegraphics[width=\hh\columnwidth]{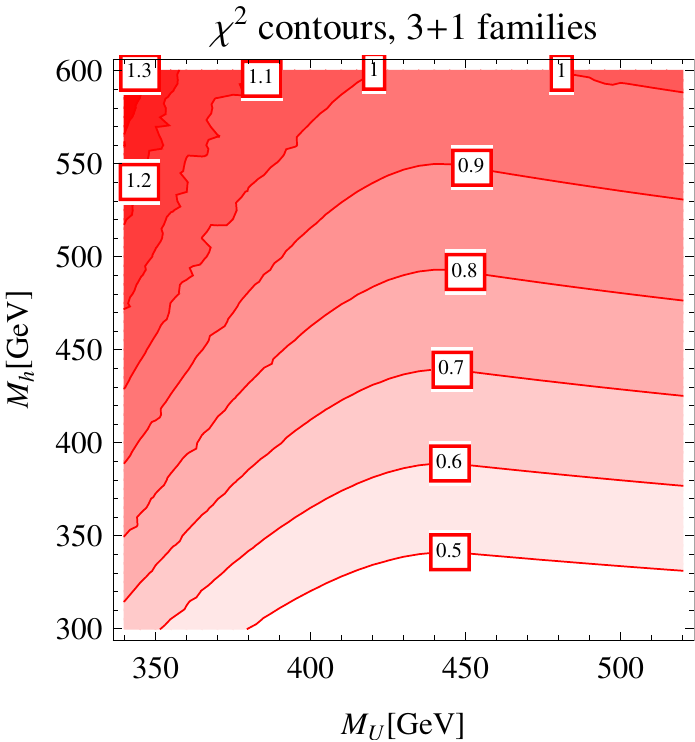}~
\includegraphics[width=\hh\columnwidth]{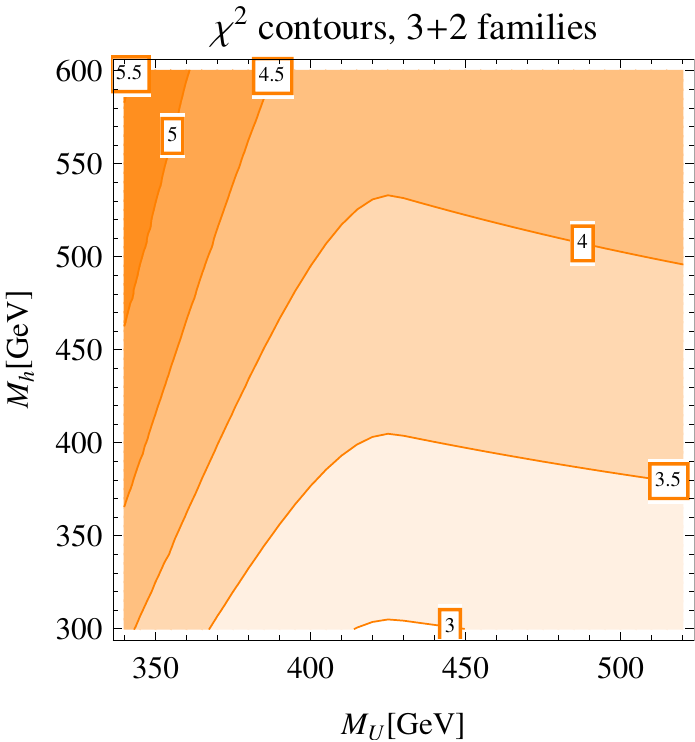}~
\includegraphics[width=\hh\columnwidth]{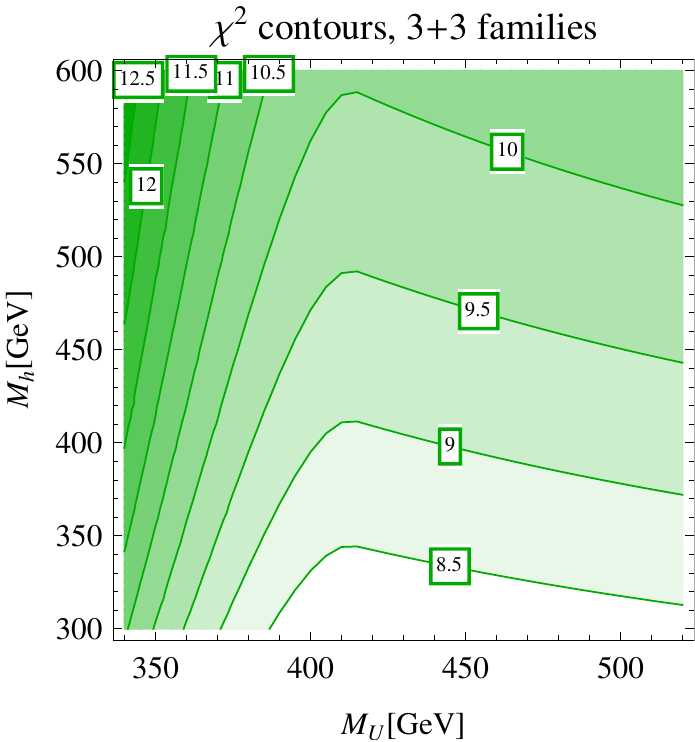}
\vspace{-2ex}%
\caption{The best $\chi^2$ contour in the $m_h$--$m_U$ parameter space, for one, two and three extra families with a second doublet. For one extra family, the SM Higgs can be either light or heavy (first and second panel) while for the case with more families, the vacuum stability constrains the Higgs to be heavy.
\label{chi2}}%
\vspace{-1.ex}
\end{figure*}

  Fig.~\ref{obliquehih} illustrates the effect of adding  an inert doublet and mirror families  on the oblique parameters  $S$, $T$ and $U$.  We show a projection in the $S-U$ plane of the 68\%, 95\% and 99\% contours from  \cite{Flacher:2008zq}, together with the value obtained for a heavy SM Higgs and how it changes with the addition first of an extra inert doublet, and then three extra families.  Sample points are given  for  a set of fermion and scalar masses that provide the best fit: $m_h=505.0\,$GeV, $m_S=50.0\,$GeV, $m_C=423.9\,$GeV, $m_A=600.0\,\GeV$, and $m_{U}=420.0\,$GeV, $m_{D}=366.2\,$GeV, $m_{N}=50.0\,$GeV, $m_E=167.0\,$GeV.

 In Fig.~\ref{chi2} we show, for one to three extra families, the contours of the best $\chi^2$ in the $m_h-m_U$ parameter space Other mass parameters are varied in the range consistent with direct search limits (see Table~\ref{expbounds} below) in order to optimize the fit, and we have taken into account the lower bound on the SM Higgs mass due to vacuum stability. Indeed, for a fourth family the Higgs can be either light ($m_h\in[114,131]\,$GeV) or heavy ($m_h>204\,\GeV$).  For more families instead, the vacuum stability bound becomes relevant: $m_h\gtrsim300\,\GeV$ and $m_h\gtrsim400\,$GeV for two and three extra families, respectively. The results, shown in Fig.~\ref{chi2}, can be summarized as follows:

\begin{itemize}
\item  The best fits are obtained when Higgs is lighter.

\item For one and two extra families one can always find solutions so that the oblique parameters are fit to within 68\% CL. 

\item In contrast, for three extra families, the SM Higgs is constrained to be heavier than 400\,GeV, so that the best $\chi^2$ turns out to be much higher, see Fig.~\ref{stability}, where we bring together the EWPT and vacuum stability and perturbativity constraints.
\end{itemize}
%

It turns out that for the mirror case the best fit scenario is very predictive regarding the mass spectrum.  First, the extra charged leptons and (especially) neutrinos are constrained to be light, while the quarks have to lie around $400\,\GeV$ and the Higgs around $500\,\GeV$ to be safe from vacuum instability (see Fig.~\ref{stability}).  Then, the scalars from the second doublet are constrained to lie in the range $250\,\GeV \lesssim m_C \lesssim 500\,\GeV$ and $m_A \gtrsim 450\,\GeV$. Also, at the best fit point, the scalar component $S$ has to be lighter than $100\,$GeV. This is only possible if $S$ has tiny (or no) mixing with the SM Higgs boson, to avoid the LEP bound on Higgs-like particles. Finally, the $\chi^2$ is also minimized when the extra doublet does not mix at all with the standard Higgs one~\cite{Martinez:2011ua}.


\subsection{An inert doublet} 

\label{sec:id}

The fact that the inert nature is favored by EWPT provides a motivation to take the lightest neutral component in the second doublet to be the dark matter candidate. Before pursuing such possibility in detail, we will define the potential and study the relevant experimental bounds on the Higgs sector.

In this scenario, the extra scalar doublet does not develop a vacuum expectation value and is not coupled to fermions~\cite{Barbieri:2006dq, LopezHonorez:2006gr, Dolle:2009fn}.  Assuming the stability of its lightest member implies an exact $Z_2$ symmetry,\footnote{Strictly speaking, this symmetry need only be approximate as long as the DM candidate is sufficiently long-lived. We discuss this issue in section III.C.} which restricts the potential to the following form:
\begin{eqnarray}
	V &= & \mu_1^2 |H|^2 + \mu_2^2 |\Phi|^2 + \lambda_1 |H|^4 + \lambda_2 |\Phi|^4
	+ \lambda_3 |H|^2 |\Phi|^2 \nonumber\\
	&&{}+ \lambda_4 |H^\dagger \Phi|^2 + 
	\frac{\lambda_5}{2} \left( (H^\dagger \Phi)^2 + \text{h.c.} \right).
\end{eqnarray}
Clearly, all terms are odd under the $Z_2$ symmetry.%

\begin{table}[b!]
   \centering
 \begin{tabular}{|l|c|}
 \hline
 LEP &  \begin{tabular}{c} $m_N >45\,\GeV$ ,  $m_E > 102.6\,\GeV$  \\$m_C > 70\,\GeV$, $m_S > 50\,\GeV$,  $m_S + m_A > M_Z$ \end{tabular}  \\
 \hline
 CDF &  $ m_U > m_D> 335\, \GeV$ , $ m_D > m_U >338\, \GeV$ \\
 \hline
 CMS &  $ m_D > 361\,\GeV$  \\
 \hline
  \end{tabular}
 \caption{Experimental bounds on extra fermion and inert doublet components, used for the fitting.}
 \label{expbounds}
\end{table}

\begin{figure*}[t]
  \centering
    \includegraphics[width=0.52\columnwidth]{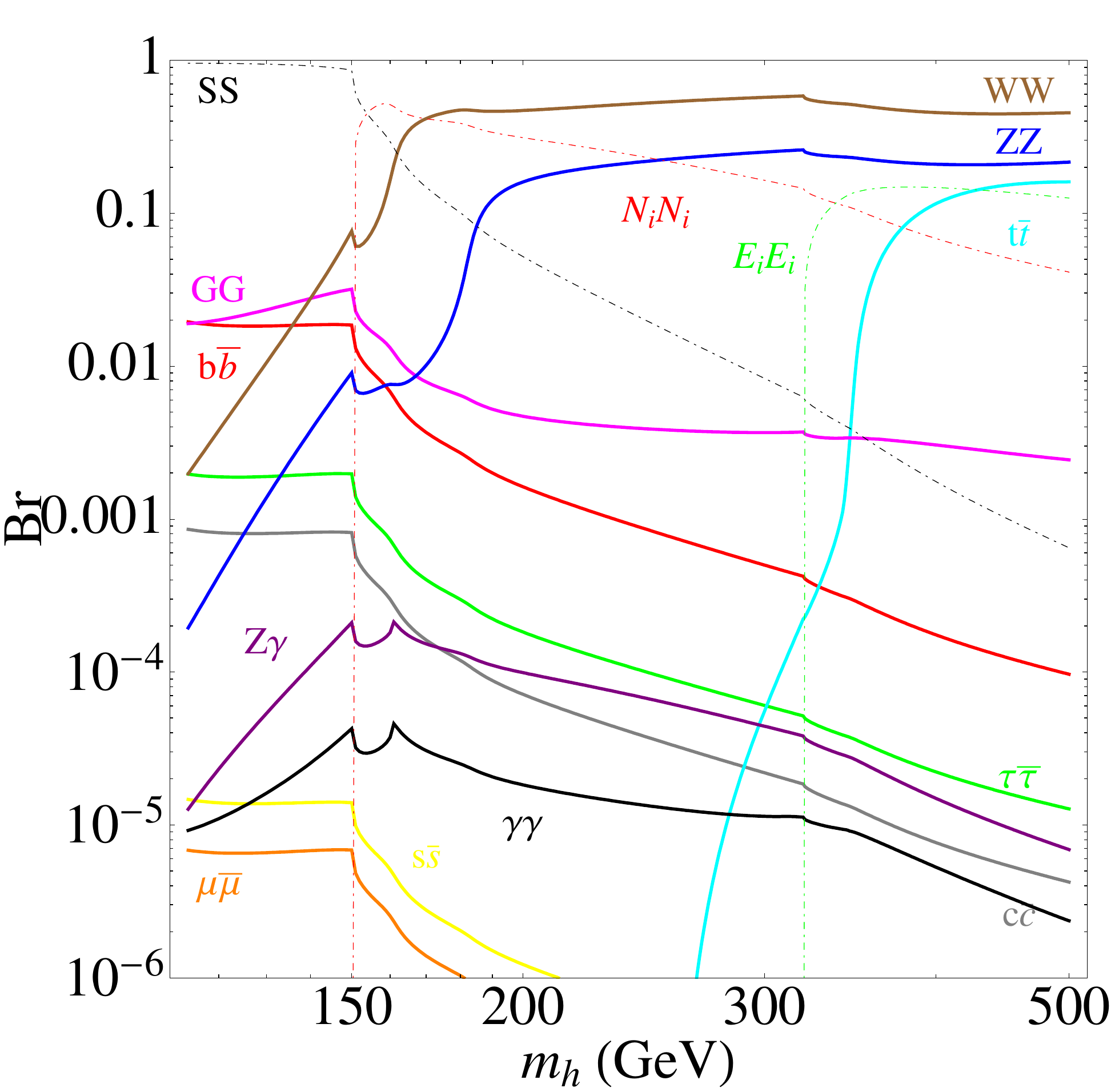}
  \includegraphics[width=0.5\columnwidth]{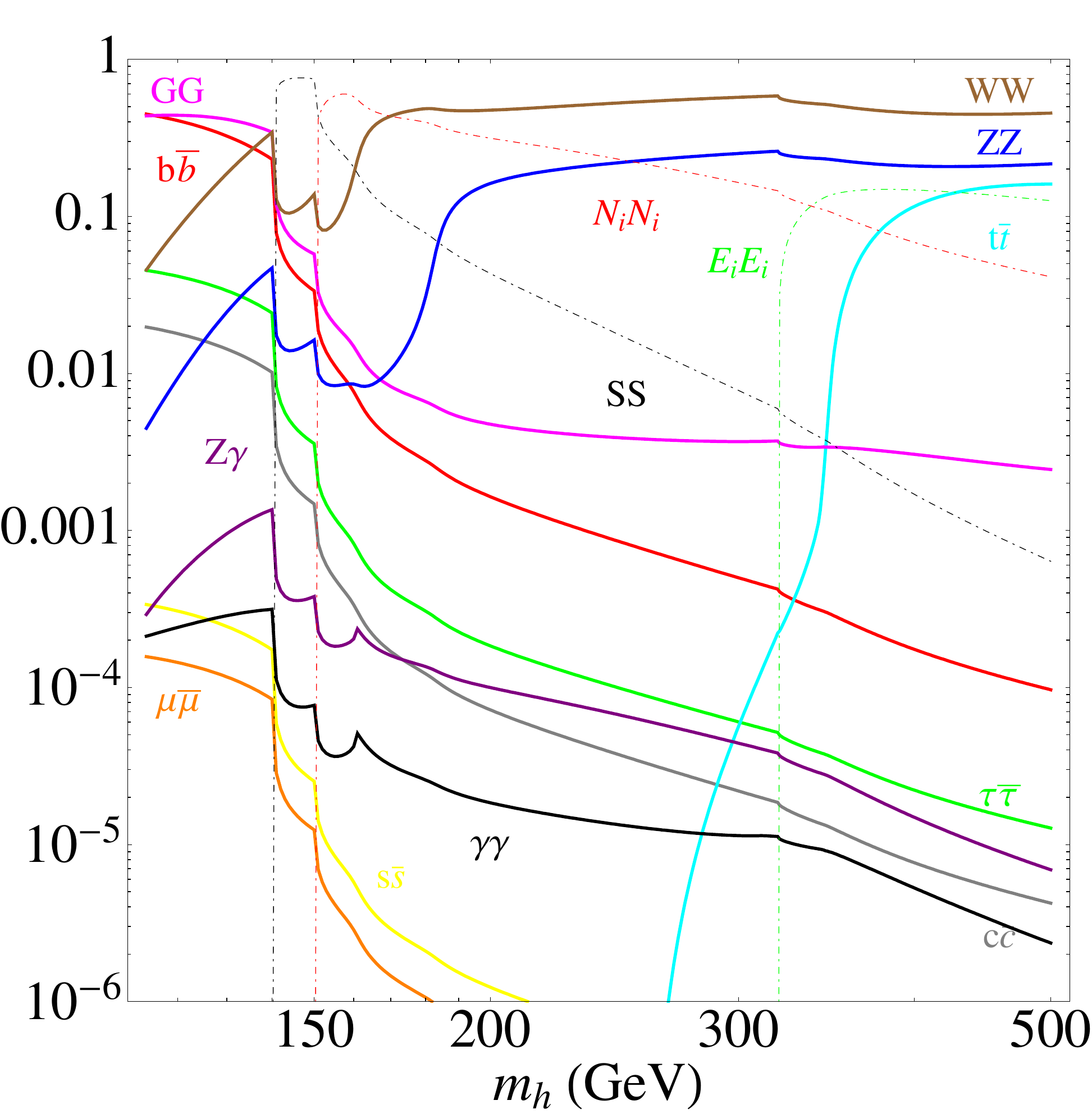}
  \includegraphics[width=0.5\columnwidth]{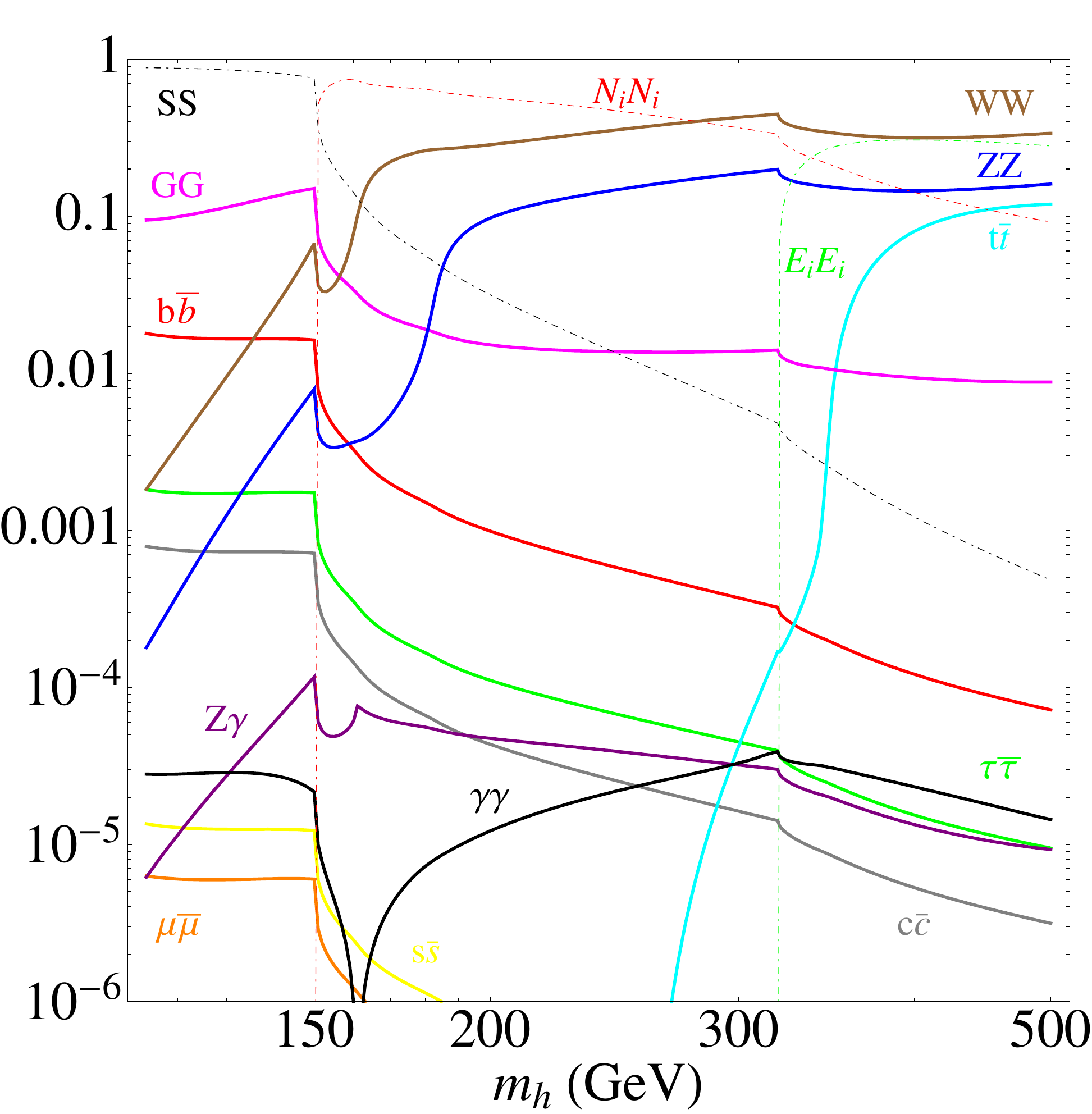}
  \includegraphics[width=0.5\columnwidth]{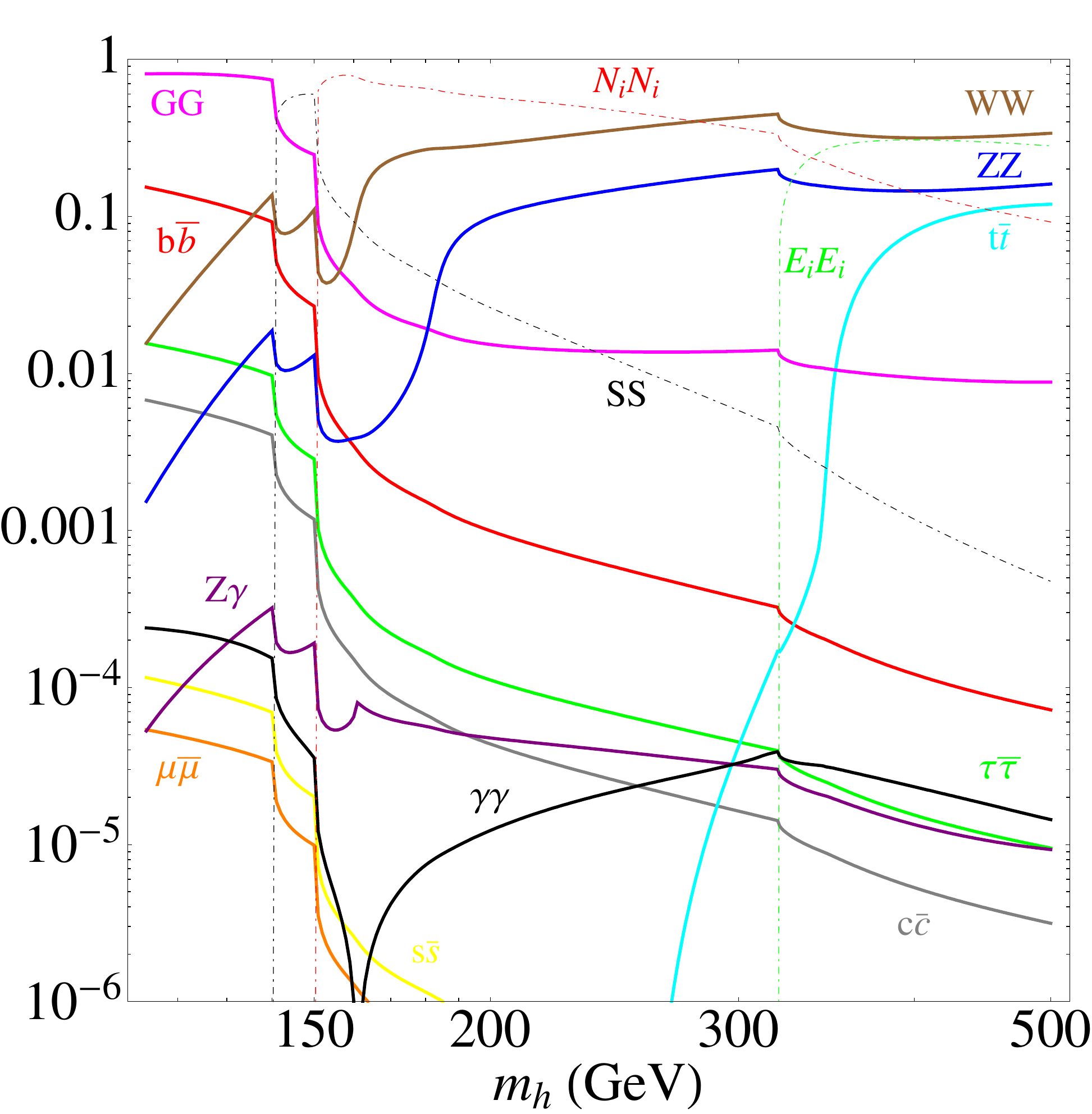}%
  \caption{Branching ratios of SM Higgs boson in the inert doublet model with a fourth family (first and second panels) and three extra or equivalently mirror families (third and fourth). In the plot, we have taken $m_S=50\,$GeV (first, third) and 70\,GeV (second, fourth), $m_{N_i}=75\,$GeV, $m_{E_i}=160\,$GeV, $m_{U_i}=m_{D_i}=350\,$GeV,
$m_C=450\,$GeV and $\lambda_L=0.3$.}%
    \label{Higgs6}
\end{figure*}

Experimental bounds on the inert doublet scalars and extra leptons derive mainly from LEP, while the extra quarks are required to be very heavy from direct searches at the Tevatron and LHC. We summarize the bounds on new fermions and inert scalars in Table~\ref{expbounds}, and refer the reader for details to~\cite{Martinez:2011ua} and references therein.


%

\subsection{Implications on the SM Higgs search} 

An extension with a second Higgs doublet and extra families changes both the production and decays of the SM Higgs boson. We calculate the branching ratios using {\tt HDECAY}~\cite{Djouadi:1997yw} as shown in Fig.~\ref{Higgs6}, with possible new decay channels into $SS$, $N\bar N$ and $E\bar E$. Meanwhile, the existence of extra quarks will enhance $H\to gg$, which dominates the branching ratios before the $SS$ decay channel opens.  Both extra quarks and leptons contribute destructively~\cite{Kribs:2007nz} with the $W$-boson to the branching ratio $H\to \gamma\gamma$. We find such destructive interference is most complete for two extra families. For one or three extra families, the suppressions of diphoton branching ratio are similar, about $0.01-0.1$ of the SM value, for a light Higgs.
We also noticed the charged scalar $C$ from the inert doublet makes a negligible contribution~\cite{Lee:2003nta} to the $H\to\gamma\gamma$ branching ratio.

On the other hand, the Higgs production cross section via gluon fusion also receives enhancement due to the presence of heavy chiral quarks.  Combining these effects, we use the most recent results on Higgs searches from D0 and CDF~\cite{Aaltonen:2010sv} to evaluate the exclusion window on the Higgs boson mass.  With the presence of a fourth family, the enhancement factor is roughly a factor of 9, for $m_h<200\,$GeV.  This has been used by~\cite{Aaltonen:2010sv} to claim the exclusion region between $131-204\,$GeV. \footnote{For critical comments about the uncertainties in this result, see~\cite{Baglio:2011wn}.}

However, this does not hold for light extra neutrino $N$ as argued in~\cite{Keung:2011zc}, because the Higgs ``invisible" decay significantly reduces the branching ratio of $WW$ channel used for the identification of the Higgs.  Similarly, if the scalar $S$ is sufficiently light (even for heavy $N$), i.e., $m_S \approx 50\,$GeV, the Tevatron exclusion window on the Higgs boson mass shrinks to $\sim150$--$200\,$GeV.  At the same time, the $H \to SS$ channel dominates for all the light Higgs mass values, as can be seen in Fig.~\ref{Higgs6}.  In any case, there is still the bound from the LEP: $m_h\gtrsim114\,$GeV~\cite{Djouadi:2005gi}.

For the mirror families case, the Higgs production cross section gets enhanced  49 times for $m_h<200\,$GeV~\cite{Djouadi:2005gi} and this factor gets reduced to as much as $\sim 20$ for heavier Higgs near the $t\bar t$ threshold~\cite{Arik:2002ci}.  In this case, the Tevatron direct search excludes the Higgs mass window between $\sim160$--$250\,\GeV$ for a light $m_S\approx 50\,$GeV and moderate $\lambda_L\approx 0.3$.  Taking a slightly heavier $m_S\approx 70\,$GeV, the exclusion window will extend to $\sim130$--$250\,\GeV$.  It is useful to recall though, that vacuum stability with mirror families excludes the light Higgs regime anyway, and further imposes the lower bound $m_h\gtrsim 400\,$GeV (Fig.~\ref{stability}).

\section{Inert Doublet as Dark matter}

\noindent As a convention, we take $S$ to be the lightest component of $\Phi$ and therefore the DM candidate (assuming $A$ to be DM is physically equivalent, since a simple redefinition interchanges them). Of course, if extra families are included, the new neutrinos are available as an additional component of DM.  This depends on the nature and mass spectrum of neutrino masses. In case they are Dirac particles, their contribution to the relic density is negligibly small, less than 0.3\%. An appreciable contribution can be obtained by a judicial choice of their Majorana masses~\cite{Keung:2011zc}.\footnote{See also~\cite{Lee:2011jk} for fourth generation RH neutrino being the DM candidate.} We do not pursue this option here, so that $S$ from the inert doublet by itself accounts for the DM.

The relevant interactions of $S$ which govern the relic density and the direct detection are its interactions with $W$ and $Z$, fixed by the
gauge group representation and the following interaction with the SM Higgs boson:
\begin{equation}
	\frac{\lambda_L v}{2} \, S^2 \, h, \quad \lambda_L = \lambda_3 + \lambda_4 + \lambda_5\,.
%
\end{equation}
Throughout the analysis, we will take $m_A=m_C\approx 450$\,GeV, a consequence of strong
hierarchy between $A$, $C$ and $S$, demanded in the case of three/mirror extra families. Therefore,
the co-annihilation effects are safely neglected.

 \subsection{Relic Density}

 To determine the relic density, we employ the {\tt MicrOMEGAs} package~\cite{Belanger:2006is}, which includes all the relevant two body annihilation final states. The relic density is constrained by the WMAP five year data~\cite{Dunkley:2008ie} to be $0.092<\Omega {\rm h}^2 <0.128$, where $\Omega$ is the dark matter density and $\rm h$ is the normalized Hubble expansion rate.

\begin{figure}[t]
  \centering
    \includegraphics[height=0.41\columnwidth]{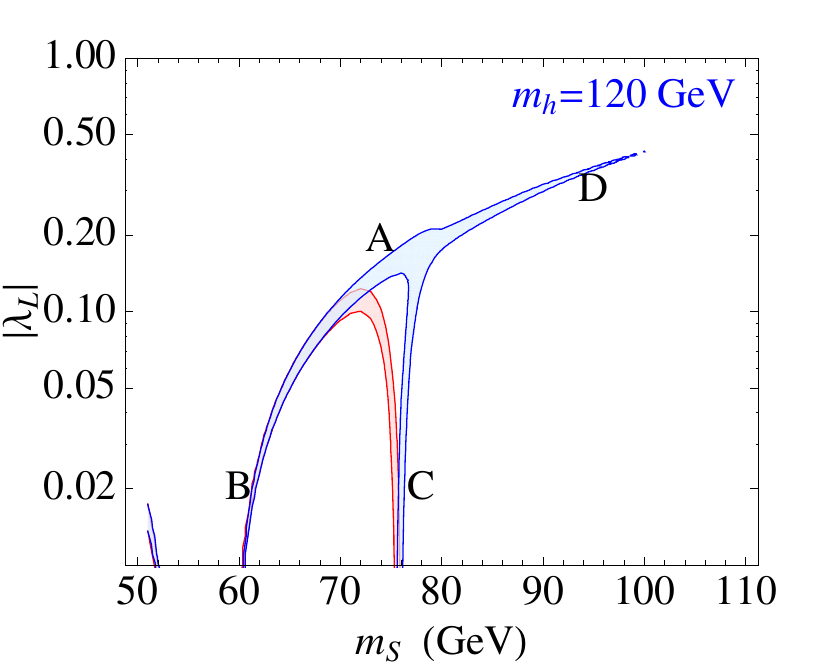} \hspace{-0.3cm}
  \includegraphics[height=0.409\columnwidth]{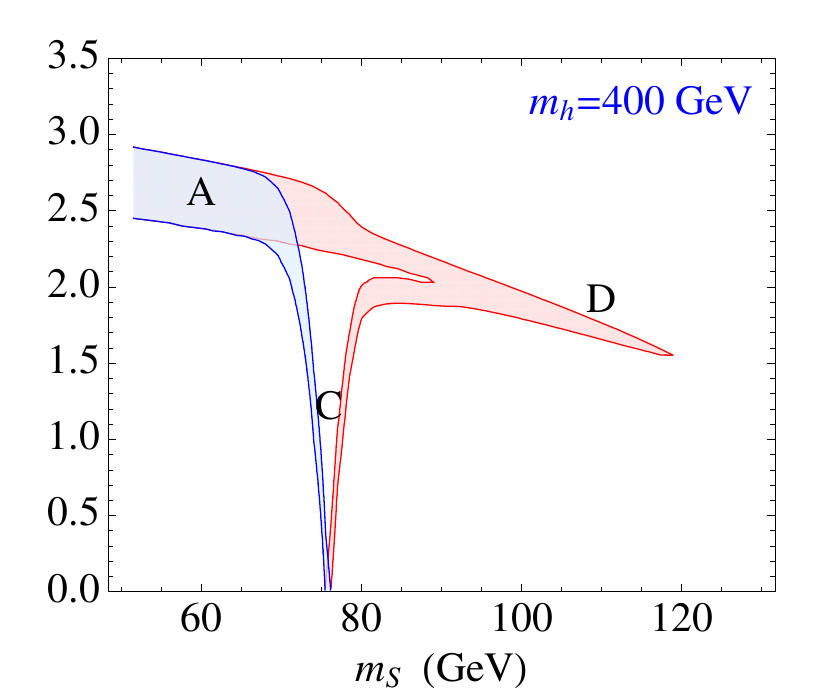}%
\vspace*{-1ex}%
  \caption{WMAP allowed parameter space in the $\lambda_L $--$m_S$ plane, for $M_h = 120$ (left) and $M_h = 400$ (right). 
   The labels of regions refer to: $SS \to b \bar b$ away from the Higgs mass pole (A);   $SS \to b \bar b$ on the Higgs mass pole (B); 
  $SS \to W W$ near the threshold (C) and $SS \to WW$ for larger $m_S$ (D). The red and blue colors stand for positive and negative $\lambda_L$, respectively.\label{animal zoo}}%
\end{figure}

The main processes controlling the thermal freeze-out of dark matter include the usual annihilation to weak gauge bosons, as well as the annihilation through the SM Higgs boson into $f\bar f$ (predominantly $b\bar b$)~\cite{Dolle:2009fn}. Thus, the relic density of the dark matter depends not only on $m_S$, but also on $m_h$ and $\lambda_L$. Roughly, the viable parameter space can be divided into the following relevant regions which are 
depicted in Fig.~\ref{animal zoo}.
 
 \begin{itemize}
 
 \item First, $S \,S \to h^* \to b \, \bar b$ (denoted A in Fig.~\ref{animal zoo}) dominates the annihilations. This can happen for a light $m_S<75\,\GeV$ (this is when the $WW$ channel takes over) and large enough $\lambda_L/m_h^2$. Alternatively, the same happens for smaller $\lambda_L/m_h^2$ but when the center of mass energy is near the Higgs pole (denoted B in Fig.~\ref{animal zoo}). The latter case corresponds to $m_S\approx \frac{1}{2}m_h$, as long as $m_h<150\,$GeV.
 
 \item Second, $S\,S\to W\,W$ dominates. This can happen either predominantly through the direct $SSWW$ coupling (denoted C), in which case in order to give the correct relic density $m_S$ is forced to lie around 75\,GeV;  or through both the direct and Higgs mediated $SSWW$ couplings (denoted D). In this case, for proper values of $\lambda_L/m_h^2$, one may obtain the correct relic density through judicious cancellation of the two contributions~\cite{LopezHonorez:2010tb} and the mass of $S$ extends from 75 to $\sim 110\,$GeV. 
 
 \end{itemize}
 
\medskip

In principle, the two body final state can be considered just as a subset of a more general annihilation channel $S \,S \to W \, W^*$.  As was noticed in~\cite{Honorez:2010re}, the three body process becomes more relevant for $m_S\lesssim75\,$GeV, when $S \,S \to b \, \bar b$ annihilation rate is low.  The three-body annihilations have not yet been included in {\tt MicrOMEGAs}, therefore the relic density $\Omega' {\rm h}^2$ provided by {\tt MicrOMEGAs} has to be rescaled to properly account for such an effect. In practice, we calculate the thermally averaged annihilation cross sections for both $S \,S \to W \, W$ and $S \,S \to W \, W^*$.  Then, the correct relic density $\Omega {\rm h}^2$ is suppressed by the factor
\begin{equation}
r=	\frac{\Omega}{\Omega'} = \frac{\langle \sigma v \rangle_{S S \to b \bar b} + 
	\langle \sigma v \rangle_{S S \to W W}}{
	\langle \sigma v \rangle_{S S \to b \bar b} + \langle \sigma v \rangle_{S S \to W W^*}}
	\, ,
\end{equation}
where the thermally averaged cross sections are evaluated at $T_f = m_S / 25$.  

 It is useful to note that the branching ratios depicted in Fig.~\ref{Higgs6} help to properly determine the SM Higgs propagator when the annihilation happens at the resonance.

\medskip

An important point to note is the possibility of annihilation of $S$ into neutrinos from extra families, which have large couplings to the Higgs boson. If such an annihilation channel is open, because of the large Yukawa couplings of $N$,
the $SSh$ coupling must be dramatically reduced in order to keep the relic density intact. In this case, the direct detection cross section is accordingly reduced, and will typically end up being below the Xenon sensitivity. 
There is also an intermediate scenario with $m_N$ just slightly below $m_S$, where the direct detection may still be possible.
We will comment on this possibility below.  Let us first focus on the scenario where all the extra neutrinos are heavier than $S$. 
%

 \begin{figure*}[t!]
  \centering
  \includegraphics[height=.7\columnwidth]{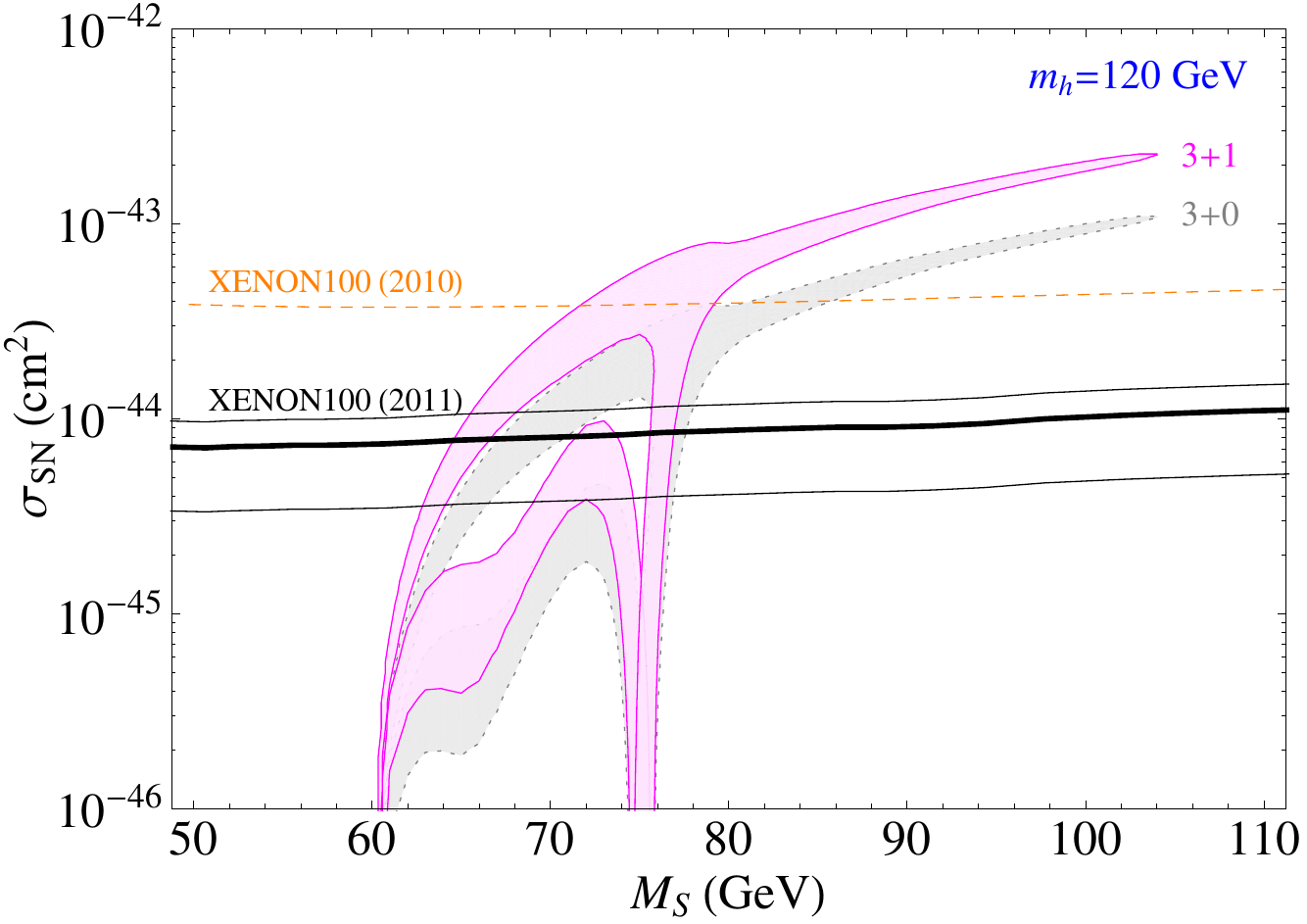} 
  \includegraphics[height=.725\columnwidth]{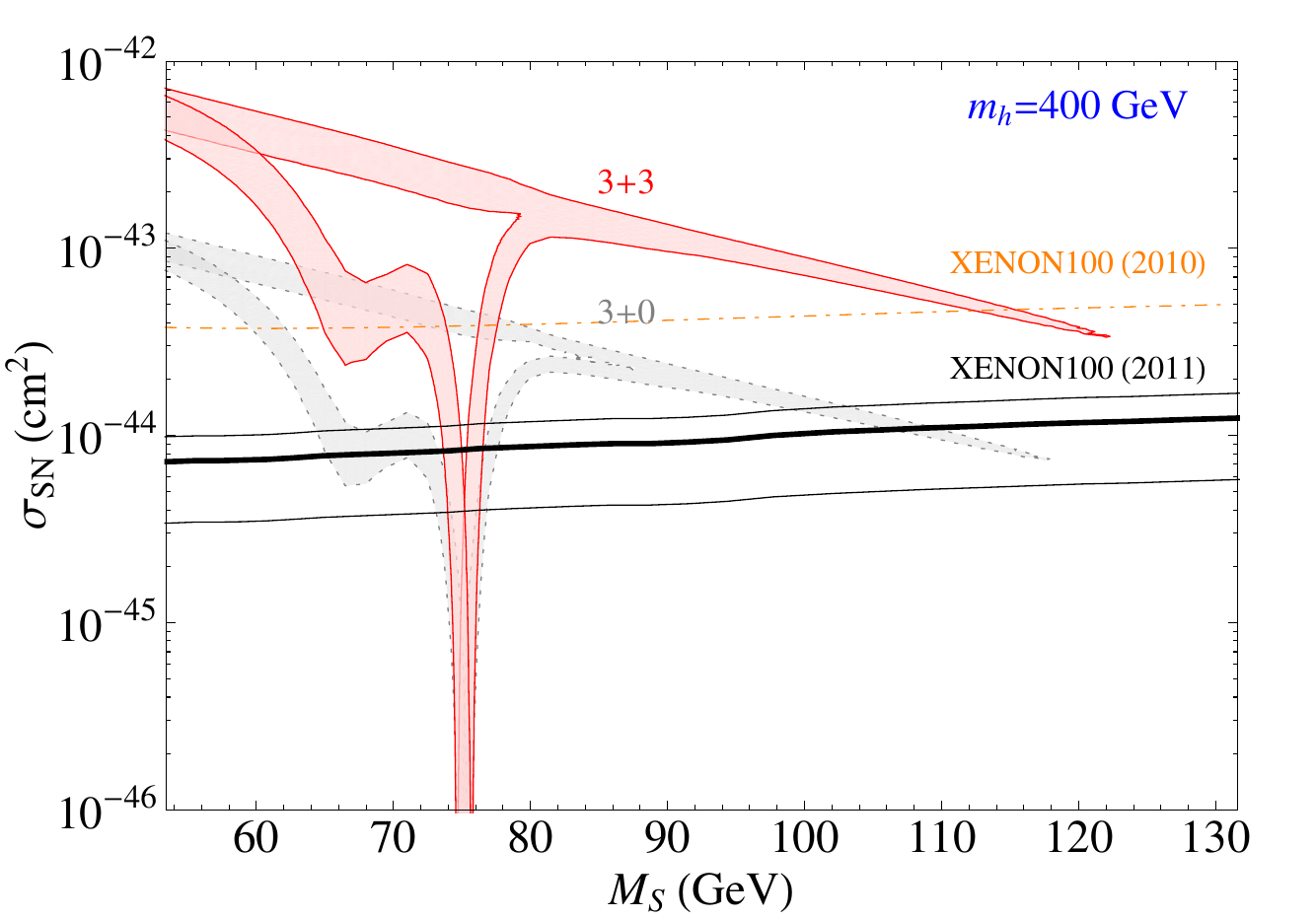}
\vspace*{-1ex}%
  \caption{ Direct detection cross section on nucleon consistent with WMAP relic density region. Left Panel: light SM Higgs regime, $m_h=120\,$GeV, with one (solid region) and zero (dotted region) extra families. Right panel: heavy SM Higgs regime, $m_h=400\,$GeV  with three (solid region) and zero (dotted region) extra families. The horizontal solid lines show the Xenon100 limits for different local dark matter densities, $\rho_\odot=0.2, 0.3, 0.6\,\GeV/\text{cm}^3$ (upper to lower solid lines). The dashed line shows the limit from the previous (2010) Xenon data release.}\label{directdetection}%
\end{figure*}

\subsection{Direct Detection with Extra Families}

Direct detection of the inert dark matter is mediated by the exchange of the SM Higgs boson with nucleons at tree level~\cite{Shifman:1978zn}. The effective matrix element for the Higgs interaction with the nucleon is~\cite{Giedt:2009mr},
\begin{eqnarray}
&&	\frac{1}{v}\langle N| \sum_q m_q \bar q q |N\rangle = \frac{m_N}{v} f(n_h) \ , \nonumber \\
&&	f(n_h) = \left( 1 + \frac{2n_h}{27} \right) ( f_{T_u}^{(N)} +  f_{T_d}^{(N)} +  f_{T_s}^{(N)}) + \frac{2n_h}{27},
\end{eqnarray}
where the sum over $q$ goes through all the quarks, $m_N$ is the nucleon mass, $n_h$ is the number of heavy quarks and $\langle N|m_q \bar q q|N\rangle \equiv m_N f_{T_q}^{(N)}$ is the nucleon sigma term for light flavors.
Clearly, the strength of the effective interaction depends on the number of heavy quarks, which contributes democratically.
In the following analysis, we use the central values of $f_{T_q}$ in~\cite{Giedt:2009mr} from the lattice results and get $f(n_h=3)=0.375$ for the SM which is close to the central value used in~\cite{Andreas:2008xy}.\footnote{Our conclusions remain the same if a relatively higher value of $f=0.467$ is used, as in the {\tt MicrOMEGAs} package. In order to be as conservative as possible, a lower value of $f$ is used throughout the paper.} The extra family extension will boost such interactions, yielding $f(5)=0.542$ for fourth family case and $f(9)=0.875$ for three extra families.
The main uncertainty in the matrix element comes from the strange quark contribution~\cite{Ellis:2008hf}. 
 The direct detection cross section is thus
\begin{equation}
	\sigma_{SN} = \frac{\lambda_L^2 f(n_h)^2}{4\pi} \frac{\mu^2 m_N^2}{m_h^4 m_S^2} \ ,
\end{equation}
where $\mu = m_S m_N/(m_S+m_N)$. In the case of mirror families there are 6 new heavy quarks and the direct detection cross section gets enhanced by a factor of 9. This facilitates the direct detection of this dark matter candidate.

It is then important to compare these predictions with the bound resulting from the recent Xenon100 released data~\cite{Aprile:2011hi}.  For a realistic comparison, one has to also take into consideration the large uncertainty in the local DM density $\rho_\odot$. This quantity, which is necessary in converting the Xenon expected rate to the excluded cross section, is inferred only from very indirect and uncertain measurements, and can at best be constrained to lie in the fairly broad range $\rho_\odot=0.4\pm 0.2\,\GeV/\text{cm}^3$~\cite{Salucci:2010qr}.  This uncertainty then shifts the bound on the cross section, and is of relevance for the model under consideration.

In Fig.~\ref{directdetection} we present the results of the comparison, for $3+1$ families (left plot) which favors light Higgs and $3+3$ families (right plot) which favors a heavy Higgs boson. These represents the main results of our work.  As it can be seen, the upper bound set by Xenon100 narrows down the allowed region for the $m_S$, even in the hypothesis of low DM density, to one or two fixed values of $m_h$.

\smallskip

\begin{itemize}
\item
Focusing first on the $3+1$ case, if the SM Higgs is light, the DM mass is practically fixed by the legs of the ``giraffe". For example, for $m_h=120\,$GeV, the mass lies in the window between $\frac{1}{2}m_h$ and $76\,$GeV. The former value corresponds to the annihilation through the SM Higgs boson to $b\bar b$ final states (with minor corrections from $WW^*$). The second value corresponds to annihilations to $WW$.  On the other hand, when the SM Higgs is heavy, the DM mass is confined to a particular value around 75\,GeV.


\item The $3+3$ case is represented by the red region in the right panel of Fig.~\ref{directdetection}.  The allowed region has only one ``leg'', because the Higgs has to be heavier than 400\,\GeV\ (for vacuum stability). Therefore, $m_S$ must lie very near 75\,GeV. Note that this value is only due to the $WW$ annihilation channel and thus it is independent from the Higgs mass. Actually, the exclusion of the rightmost part (in contrary to zero extra family case) is quite insensitive to the particular choice of Higgs mass (as explained for case D in figure~\ref{animal zoo}).


\end{itemize}

The Xenon collaboration also claimed a mild positive evidence of dark matter, whose cross section is just below the current bound. If true, it can be easily accommodated for the values of $m_S$ discussed above.

\medskip

Let us finally comment on the other possible scenario mentioned above, where some of the heavy neutrinos are lighter than $S$, allowing the annihilation $SS\to NN$. In this case, to maintain the correct relic density, the $hSS$ coupling $\lambda_L$ is reduced by a factor of at most $\sim 1/10$ and the direct detection regions shown in Fig.~\ref{directdetection} are shifted downwards by $10^{-2}$. This reduces the predictivity of this scenario in terms of the $S$ mass, but implies in turn that $m_N$ lies in a narrow region, $[45\,\GeV,m_S]$, which is important for the detection of $N$ at colliders.  From the point of view of EWPT, both scenarios with either lighter or heavier than $N$ are equally allowed.

\subsection{Indirect Detection with Gamma-ray Line}

As we saw above, the 100 live-day Xenon100 results restrict the DM mass to a narrow region, especially in the presence of mirror families. The main annihilation channel during freeze-out is to gauge bosons, while today in the galactic halo since the temperature is low, the annihilation through the SM Higgs to $b\bar b$ could be important. A spectacular signature of the inert doublet DM would be the observation of a monochromatic gamma-ray line from its annihilation in our galactic halo. This could serve as a promising signal of indirect detection to determine the mass of the DM.  In this model, such process goes through a dimension six operator $SSF_{\mu\nu}F^{\mu\nu}$ with a loop suppression (mainly $W$-loop).  The DM initial states $SS$ can either couple to the $W$-loop directly, or through the SM Higgs to both $W$ and heavy fermion loops.  In fact, the associated loop functions are the same as those in the $h\to\gamma\gamma$ process.  The implication of Xenon100 is the suppression of the $SSh$ coupling and thus the Higgs mediated annihilation to two photon, which eliminates the possibility of any destructive cancellation. Therefore, there is a robust prediction of a lower bound on the gamma-ray line flux.

\begin{figure}[t]
  \vspace*{-1ex}%
  \centering
  \includegraphics[width=1.00\columnwidth]{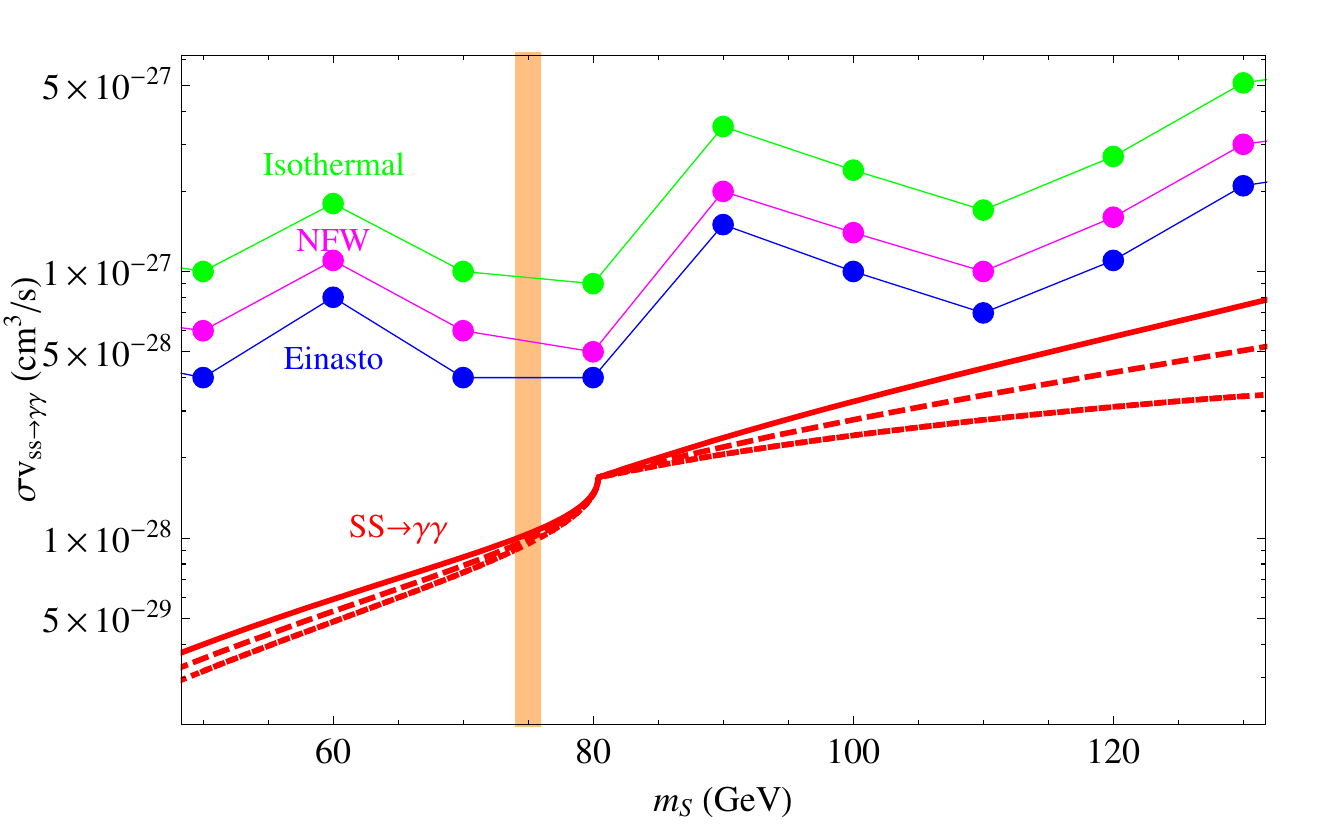}
  \vspace*{-2ex}%
  \caption{Model prediction (red curves) and Fermi LAT constraints (blue, magenta, green) assuming different DM halo density distributions on the DM annihilation cross section to two monochromatic photons in the galaxy. We have taken $m_h=400\,$GeV. The solid, dashed, dotted curves correspond to $\lambda_L=-0.5, 0.01, 0.5$ respectively. The orange shaded region is the range of DM mass consistent with relic density and direct detection.}\label{indirectdetection}%
\end{figure}

The Fermi LAT experiment has put constraints on the DM annihilation into gamma-ray lines between the energy 30--200\,GeV~\cite{Abdo:2010nc}. In Fig.~\ref{indirectdetection}, we show the cross section as a function of the DM mass for $m_h=400\,$GeV and different values of $\lambda_L$, as well as the experimental bound assuming different halo density functions.  For the mirror case where the DM mass is restricted to 74--76\,GeV by WMAP and Xenon100, we find the predicted annihilation cross section $\sigma v(SS\to \gamma\gamma)$ lies only a factor of 4--5 below the current Fermi LAT bound. If the future data release can further push the limit down by one order of magnitude, one will be able to verify or exclude the possibility of the inert doublet being the DM candidate.

\smallskip

{\em Decaying DM: approximate $Z_2$ symmetry?} Up to now, as in previous studies we have assumed the dark matter to be absolutely stable. If it were to be so, it would imply the existence of an exact $Z_2$ symmetry.
Observationally, dark matter does not have to be absolutely stable, and therefore one should be open minded to consider the possibility that the $Z_2$ symmetry is only approximate.  The point is, that approximate global symmetries are equally useful in guaranteeing the naturalness of small couplings as the unbroken one. This is the essential criterion of naturalness.
Needless to say, an unstable DM still has to be cosmologically long lived.  A decaying scalar dark matter 
could also lead to mono-chromatic gamma-rays carrying energy equal to half of its mass.  This process could go through the effective dimension five operator $({\epsilon}/{v})S F^{\mu\nu}F_{\mu\nu}$, where $\epsilon$ breaks the $Z_2$ symmetry explicitly. Fermi LAT imposes an stringent upper bound~$\epsilon\lesssim 10^{-26}$.

\section{LHC prospects}

\noindent Finally, we comment on the LHC prospect of discovering or falsifying this theoretical setup. 

\smallskip

{\em Heavy quark states.} The most obvious way to verify or falsify the above framework is to search the heavy quarks from extra families. Being colored states, they have large cross sections at hadron colliders. As mentioned in Sec.~II, the current limits on the heavy quarks are around $350\,$GeV, mainly from Tevatron. With higher energy and luminosity, LHC can soon push the mass limits into the non-perturbative regime, where new bound states could emerge whose properties largely depend on the corresponding Yukawa couplings~\cite{Ishiwata:2011ny}. 

\smallskip

{\em Inert dark matter signatures.}  The signatures of the inert doublet model have been studied in~\cite{Dolle:2009ft, Kadastik:2009gx}, focusing on the multi-lepton final states. One should keep in mind that a typical mass spectrum of the inert doublet is quite hierarchical in the mirror scenario under consideration.  In particular, we find $A$ to be the heaviest inert scalar $450\,\GeV<M_A<600\,\GeV$, the mass of $C$ lies in the intermediate range $250\,\GeV<m_C<500\,\GeV$ and $S$ is very light $50\,\GeV<m_S<150\,\GeV$.
The resulting signatures after their pair production at the LHC differ slightly from the previous analyses, due to the large mass hierarchy and potential cascade decays of both $A$ and $C$.  Therefore, the final state leptons and jets possess large transverse momentum. Meanwhile, due to the fact that any such final state always contains a pair of $S$, the resulting missing energy will also be typically larger than $100\,$GeV. These characteristics can be fully tested when the energy of LHC reaches 14\,TeV.

\smallskip

{\em Are mirror neutrinos Majorana?} A priori, just as in the SM, we cannot know the nature of neutrinos. The dominant view today is the Majorana picture which, if true, would have particularly exciting consequences for the neutrinos belonging to extra generations. Particularly interesting is the mirror case, which forces the three mirror neutrinos to be heavy neutral leptons with masses around 50-100 GeV. They could even be the source~\cite{Hung:2006ap} of the seesaw mechanism in which case, the mirror and ordinary families would be forced to mix by a tiny amount. Although this is not mandatory, these mixings are plausible and are naturally small enough (technically, the mirror symmetry preserves their smallness) to evade the $Z$ width constraint.  In all honesty, this appealing, simple and testable seesaw picture may not be very convincing. After all, one needs new physics to generate the Majorana masses of mirror neutrinos and there is no reason that the new physics is not generating Majorana masses for ordinary neutrinos, too.

What about the seesaw paradigm, with only one or two extra families? The former case is immediately ruled out since one predicts only one massive ordinary neutrino.  In the latter case, one has an interesting prediction of maximally hierarchical neutrinos, since only two of them are massive. This fits nicely with cosmological considerations, which keep lowering the sum of light neutrino masses~\cite{Hannestad:2010yi}.  Moreover, the decays of the heavy extra neutrinos $N$ are governed by the Dirac mass terms, which are functions of the leptonic mixing matrix, the masses of $N$'s and only one complex parameter~\cite{Ibarra:2003up}. This case can definitely be tested by measuring different flavor combinations of the final dilepton final states, similar to the minimal case of type I+III seesaw~\cite{Bajc:2007zf}. This could be an example of a testable seesaw mechanism at the LHC. If one gives up the dark matter candidate, EWPT work in the minimal setup with only the standard Higgs doublet, in which case the masses of $N$'s lie again between $50\,$GeV and $150\,$GeV.

In any case, irrespective of the seesaw, it is worth considering mirror symmetry not to be exact.  Once $N$ and $E$ are produced pairwise, their decay can lead to the interesting two leptons and six jets events with no missing energy~\cite{Carpenter:2010bs}.  If $N$ is of Majorana type, the characteristic feature is the equal decay rate in leptons and antileptons~\cite{Keung:1983uu}.

\section{Conclusion and Outlook}

\noindent In recent years, the inert scalar doublet model has become one of the popular extensions of the SM whose lightest component can play a role of the DM. Whereas its simplicity may be appealing, the inert nature is postulated by hand, which makes it more a model {\it for} rather than {\it of} dark matter.  On the other hand, the inert nature is a natural scenario with the existence of mirror families, due to the electroweak precision constraints~\cite{Martinez:2011ua}.  Mirror fermions have been suggested more than 50 years ago, as a way of restoring parity, and are well motivated by a number of respected theoretical frameworks: KK compactification, $N=2$ supersymmetry, family unification based on large orthogonal groups and some unified models of gravity.
  
This has encouraged us to carefully study the issue of dark matter in the context of the inert scalar doublet and mirror fermions. Since nothing in particular depends on the chirality of extra families, we have broadened our study by including the cases of only one and two extra families. The fourth generation has recently been the focus of a large body of research and as such deserves a special merit, in spite of the scalar's inertness not being called for. The case of two extra families does not possess any special features and thus we only commented on it in passing only. We now summarize the essential features case by case.
  
{\em One extra family.} It is not surprising that this case passes the EWPT since it works even with only one Higgs doublet as in the SM. For sufficiently light $S$ and/or extra neutrino, the Higgs mass is excluded from the window 150--200\,GeV by Tevatron data; otherwise the exclusion window is larger, 131--204\,GeV~\cite{Aaltonen:2010sv}.  As far as the direct DM detection is concerned, we find that the Xenon100 result restricts the DM mass to lie in the window $\frac{1}{2}m_h$--$76\,$GeV if the SM Higgs is light, and almost fixed at \hbox{74--76}\,GeV if the Higgs is heavy.
 
{\em More extra families.} Let us recall that since the extra quarks have to lie above the direct limits, their large Yukawa couplings rule out the light Higgs window because of vacuum stability~\cite{Martinez:2011ua}. It is then enough to consider a heavy Higgs $m_h\gtrsim300$--$400\,\GeV$. This is just above the Tevatron exclusion region which extends, in the case of three families, up to $m_h\gtrsim 240\,\GeV$.  In terms of direct DM detection, the Xenon100 experiment together with the WMAP relic density constraint makes this scenario very predictive. In fact, the hadronic uncertainty in the $h$-nucleon coupling barely plays a role here, and the DM mass turns out to lie necessarily at 74--76\,GeV.  This could lead also to a characteristic signature in indirect DM search, in terms of the spectrum of particles resulting from both annihilation or decay from galactic haloes.

\section*{Acknowledgments}

\noindent The authors would like to thank BIAS for the inspiring and conductive research atmosphere.  YZ would like to thank Kev Abazajian, Kaustubh Agashe, Zackaria Chacko, Rabi Mohapatra and Raman Sundrum for fruitful comments, and acknowledge hospitality from theoretical hadronic physics and elementary particles groups at University of Maryland during the final stage of this work.

\appendix

\begin{widetext}

\section{Explicit formul\ae\ for electroweak oblique parameters: Higgs sector}
\label{app}

\noindent In the calculation of electroweak oblique parameters $S$, $T$ and $U$ for the Higgs sector, we apply the generic formulas given in~\cite{Grimus:2008nb} to the case with two Higgs doublets. The SM Higgs mass is denoted as $m_h$, while the reference point mass (corresponding to $S=T=U=0$) is $m_r$, which is taken
to be 120 GeV throughout the paper.

\bea
  S_H &=& \frac{1}{24 \pi} \left\{  (2 s_W^2 -1)^2 \, G(m_C^2,m_C^2,m_Z^2)  +  \ln\left( \frac{m_A^2 m_h^2 m_S^2}{m_C^4 m_r^2}  \right)  \right . + \sin\theta^2 \,  G(m_A^2,m_h^2,m_Z^2) 
    + \cos\theta^2 \,  G(m_A^2,m_S^2,m_Z^2)  \nonumber \\ 
 & & \left.     +  \sin\theta^2 \,  \hat G(m_S^2,m_Z^2)  +   \cos\theta^2 \,  \hat G(m_h^2,m_Z^2) -   \hat G(m_r^2,m_Z^2)
    \right\} \ ,
  \eea 
  
 \bea
 T_H &=& \frac{1}{16 \pi M_W^2 s_W^2} \left\{  
 F(m_C^2, m_A^2) 
  \right.  +  
  \sin\theta^2 \left[ F(m_C^2, m_h^2) -  F(m_A^2, m_h^2) \right]  
+
  \cos\theta^2 \left[ F(m_C^2, m_S^2) -  F(m_A^2, m_S^2) \right]   
   \nonumber \\ 
& & 
+ 3 \sin\theta^2\left[ F(m_Z^2,m_S^2 ) -  F(m_W^2,m_S^2 ) \right]   + 3 \cos\theta^2\left[ F(m_Z^2,m_h^2 ) -  F(m_W^2,m_h^2 ) \right]   
\left.
- 3 \left[ F(m_Z^2,m_r^2 ) -  F(m_W^2,m_r^2 ) \right]
 \right\} \ ,\nonumber \\
 \eea
 
 \bea
 U_H& =&   \frac{1}{24 \pi}   \left\{   - (2 s_W^2 -1)^2 \, G(m_C^2,m_C^2,m_Z^2)  + G(m_C^2,m_A^2,m_W^2)   \right.  + \sin\theta^2 \,  \left[ G(m_C^2,m_h^2,m_W^2)  -  G(m_A^2,m_h^2,m_Z^2) \right] 
  \nonumber \\
  & &  + \cos\theta^2 \,  \left[ G(m_C^2,m_S^2,m_W^2)  -  G(m_A^2,m_S^2,m_Z^2) \right]  + \sin\theta^2 \,  \left[ \hat G(m_S^2,m_W^2)  - \hat G(m_S^2,m_Z^2) \right] 
   \nonumber \\
  & &  + \cos\theta^2 \, \left[ \hat G(m_h^2,m_W^2)  - \hat G(m_h^2,m_Z^2) \right]  
      \left. 
   - \left[ \hat G(m_r^2,m_W^2)  - \hat G(m_r^2,m_Z^2) \right]  \right\} \ ,
 \eea
where
\bea
G(x_1,x_2,x_3)  &=& -\frac{16}{3} + 5 \frac{x_1 + x_2}{x_3} - 2 \frac{(x_1 - x_2)^2}{x_3^2}  + \Delta\frac{f(x_1,x_2,x_3)}{x_3^3}   + \frac{3}{x_3}\left( \frac{x_1^2 + x_2^2}{x_1 - x_2}- \frac{x_1^2 - x_2^2}{x_3}  + \frac{(x_1 - x_ 2)^3}{3 x_3^2} \right)\ln\frac{x_1}{x_2} \ , \nonumber \\
\eea

\beq
f(x_1,x_2,x_3) = \left \{ \begin{array}{l r} 
\sqrt{\Delta} \ln \left|\frac{ x_1 + x_2 - x_3 - \sqrt{\Delta} }{x_1 + x_2 - x_3 + \sqrt{\Delta}}\right|  \; & \;  \Delta > 0 \\
2 \sqrt{-\Delta}
\left[\arctan\left( \frac{x_1-x_2+x_3}{\sqrt{-\Delta}}  \right)- \arctan\left( \frac{x_1-x_2-x_3}{\sqrt{-\Delta}} \right) \right]\; & \;  \Delta < 0 \\
\end{array} \right.  \ ,
\eeq
\beq
\Delta = x_3^2 - 2 x_ 3 (x_1 + x_2) + (x_1 - x_2) ^2 \, , 
\eeq

\beq
F(x_1,x_2) = \frac{x_1 + x_2}{2} - \frac{x_1 \, x_2 }{x_1 - x_2} \ln\frac{x_1}{x_2} \ ,
\eeq

\beq
\hat G (x_1,x_2)= G(x_1,x_2,x_2) + 12 \left[ -2 + \left( \frac{x_1 - x_2}{x_2} - \frac{x_1 + x_2}{x_1 - x_2}\right)\ln\frac{x_1}{x_2}  +  \frac{f(x_1,x_2,x_2)}{x_2} \right ] \ .
\eeq

\end{widetext}

\end{document}